\documentclass[pra,aps,groupedaddress,twocolumn,floatfix,nofootinbib]{revtex4}
\usepackage{graphicx}
\usepackage{soul}
\usepackage{float}
\usepackage{centernot}

\newcommand{\beq}{\begin{equation}}
\newcommand{\eeq}{\end{equation}}
\newcommand{\beqa}{\begin{eqnarray}}
\newcommand{\eeqa}{\end{eqnarray}}

\def\half{\frac{1}{2}}
\def\opone{\leavevmode\hbox{\small1\normalsize\kern-.33em1}}

\begin{document}

\title{Indeterminism in Physics and Intuitionistic Mathematics}
\author{Nicolas Gisin \\
\it \small   Group of Applied Physics, University of Geneva, 1211 Geneva 4,    Switzerland}

\date{\small \today}
\begin{abstract}
Most physics theories are deterministic, with the notable exception of quantum mechanics which, however, comes plagued by the so-called measurement problem. This state of affairs might well be due to the inability of standard mathematics to ``speak" of indeterminism, its inability to present us a worldview in which new information is created as time passes. In such a case, scientific determinism would only be an illusion due to the timeless mathematical language scientists use. To investigate this possibility it is necessary to develop an alternative mathematical language that is both powerful enough to allow scientists to compute predictions and compatible with indeterminism and the passage of time. We argue that intuitionistic mathematics provides such a language and we illustrate it in simple terms.
\end{abstract}
\maketitle

\section{Introduction}\label{intro}
Physicists are not used to thinking of the world as indeterminate and its evolution as indeterministic. Newton's equations, like Maxwell's and Schr\"odinger's equations, are (partial) differential equations describing the continuous evolution of the initial condition as a function of a parameter identified with time. Some view this state of affairs as the paradigmatic signature of scientific rigor. Accordingly, indeterminism could only be a weakness due, for instance, to an incomplete description of the situation.

There are essentially two kinds of common objections to the above determinsitic worldview. First, not all evolution equations one encounters in physics have a unique solution for all initial conditions. Even the classical Hamiltonian equations may fail to satisfy the Lipschitz conditions and thus allow for several solutions, possibly even a continuous infinity of solutions, as is the case of Norton's dome \cite{NortonDome}. These cases are fairly contrived and exceptional, but still worth remembering when colleagues claim that determinism is obvious. Second, there is quantum physics, which is generally presented as intrinsically indeterministic. Actually, one can even prove the existence of quantum randomness from two highly plausible assumptions: physical distances exist (nothing jumps arbitrarily fast from here to there\footnote{In other words, no influence ``propagates" at infinite speed; this excludes, among others, Bohmian mechanics \cite{Bohm52,BellBohm,DurrTeufel}.}), and no super-determinism (no combination of determinism and conspiracy, i.e.~there are de facto independent processes) \cite{Pironio10, Acin16,BrunnerRMP14,GisinQchance14}. But this quantum indeterminism is immediately associated with difficulties. These come either under the name of the measurement problem \cite{MaudlinMeasProb}, or claims of the incompleteness of quantum theory that should be complemented by additional variables, as in Bohmian mechanics \cite{Bohm52,BellBohm,DurrTeufel}, or by the radical many-worlds view \cite{Kent09}. I address quantum indeterminacy in appendix \ref{Qindeterminacy}. But allow me to straightforwardly continue with my main motivation.

I have always been amazed by the huge difficulties that my fellow physicists seem to encounter when contemplating indeterminism and the highly sophisticated circumlocutions they are ready to swallow to avoid the straightforward conclusion that physics does not necessarily present us a deterministic worldview. But even more surprising to me is the attitude of most philosophers, especially philosophers of science. Indeed, most philosophers who care about quantum physics adopt either Bohmian mechanics \cite{Bohm52,BellBohm,DurrTeufel} (become a Bohmian, as one says) or the many-worlds interpretation \cite{Kent09}. Apparently, to most of them adding inaccessible particle positions or inaccessible parallel worlds to their ontology is a reasonable price to pay in order to avoid indeterminism. The central claim of Bohmians is that they circumvent the quantum measurement problem. The latter has two sides. First, quantum theory is silent about when potentialities become actual. Second, quantum randomness seems to emerge from nowhere (from outside space-time, as I sometimes wrote) and does thus not satisfy Leibniz's principle of sufficient reason. Admittedly, the first point is serious. However, the second one is unfair: in all fundamentally indeterministic theories there must be events that happen - and thus information that gets created - although their happenings were not necessary. Spontaneous quantum collapse models are good examples of such consistent indeterministic theories\footnote{Moreover, several of such collapse models also answer the first side of the measurement problem.} \cite{GRW,belljumps,NGHPA89,RMPcollapseModels}. Consequently, since indeterminism is precisely the view that time passes and creates new information, one can't argue against indeterminism by merely asserting that the creation of new information is a priori impossible.

A very different reaction to these difficulties is illustrated by Yuval Dolev's claim that {\it tense and passage are not, never were, and probably cannot be part of physics and its language} \cite{DolevSilence}.

But is indeterminism truly that difficult to conceptualize and contemplate \cite{FlavioNG19}? Isn't indeterminism pervasive in our lives, and shouldn't physics not only lead to fascinating technologies and sophisticated theories, but also tell stories of {\it how nature does it} in a language that allows humans to gain intuitive understanding? If not, how could our intuition develop? And, is there deep understanding without intuition? Shouldn't one reply to Dolev by adapting the mathematical language used by physics to make it compatible with indeterminism? Shouldn't one paraphrase Rabelais and state that ``Science without time is but the ruin of intelligibility" \cite{NGNaturePhysComment20}?

It is worth trying to accept indeterminism at face value and see where this leads us. Accordingly, let's assume that nature has the power to continually and spontaneously produce randomness in the form of entirely new information (though information without meaning). For simplicity, and because we need to start somewhere, let's assume this power expresses itself by a {\it natural random process} that continually produces random bits at discrete instants of time. Continuous random processes and larger alphabets would also work, but let's start simple, i.e.~with binary information in the form of bits produced at discrete time steps, which I name ``time-instants". Is it truly impossible to think of such a fundamental {\it Natural Random Process} (NRP)? Not a human-made one, but - again - as a power of nature. Isn't that very natural if one wants to contemplate the possibility that our world is intrinsically indeterministic? Can one start anywhere else than by assuming that our world (nature) has the power to produce random bits? Let me emphasize that I am not claiming that physics is indeterministic; more modestly, I am interested in showing that our best physics is compatible with indeterminism\footnote{Rietdijk and Putnam have argued that (special) relativity is incompatible with indeterministim \cite{Rietdijk66,Rietdijk76,Putnam67}. However, I believe that Stein and Savitt's responses are convincing \cite{Stein91,Savitt09}. Einstein is often quoted for his claim that ``God doesn't play dice", but his position was actually quite more subtle \cite{FlavioEinsteinRealism}.}.

Let us contrast the above assumption of the existence of a natural random process with the common assumption that real numbers faithfully describe our world, in particular that the positions of elementary particles (or their centers of mass) in classical and in Bohmian mechanics are faithfully described by mathematical real numbers \cite{NGrealNb}. Typical real numbers contain infinite (Shannon) information. Hence, the ``real number assumption" is far from cheap. It assumes that ungraspable infinite quantities are at the basis of our physics. This is certainly not an obvious and clean assumption\footnote{For example, when one says ``let $x_0$ be the initial condition", one is effectively saying ``let $x_0$ denote an inaccessible infinite amount of information about the initial condition".}. 

The infinite information necessary to describe a typical real number is clearly seen in its binary (or decimal) expansion: the series of bits (digits) never ends and - typically - has no structure whatsoever. Admittedly, there are exceptional numbers for which the bits have a great deal of structure, like rational and computable numbers, i.e.~numbers $x$ for which there is an algorithm that generates a series of approximations\footnote{The bits of computable numbers may look random, but are fully determined by a (finite) algorithm. Interestingly, there are relatively simple formulas that allow one to compute any bit of $\pi$ without the need to first compute all previous bits (see subsection \ref{CN} and \cite{piDigits}). This clearly illustrates that these bits are not random but determined.} $x_n$ such that $|x-x_n|\leq 2^{-n}$ for all positive integers $n$. Although essentially all numbers one encounters are exceptional (they have names and there are only countably many names), the vast majority of real numbers are not computable; their bits (and digits) are truly random, as random as the outcomes of quantum measurements, and are of maximal Kolmogorov complexity. This randomness is crucial for Bohmian mechanics to reproduce quantum statistics and for classical dynamical systems to produce chaos\footnote{Recall that almost all classical dynamical systems are non-integrable, hence their solutions are hyper-sensitive to the initial condition, i.e.~are chaotic.}. The latter is often termed ``deterministic chaos", though is it based on real numbers whose decimals are random \cite{Chaitin1,Chaitin2}.

In \cite{NGrealNb} I argued that one should avoid infinite information densities, i.e.~that a finite volume of space can contain at most a finite amount of information (see also \cite{Drossel,DowekRealNb13,NPcommentEllis}). Then I concluded that the mathematical real numbers should be replaced by finite-information numbers and that the usual real numbers would be better named ``random numbers". In this way, whenever the equations have an analytic solution and the initial conditions and the time $t$ are given by finite-information numbers, then the state at time $t$ is also given by finite-information numbers. And if the solution is numerical, then obviously it is given by finite-information numbers. If, on the other side, the dynamical system is chaotic, then the finite-information number encoding the initial condition must be complemented by fresh random bits as time passes, which is equivalent in practice to the randomness hidden in the real/random numbers that usually describes the initial conditions.

Consequently, intrinsic randomness is already hidden in our archetype of a deterministic physics theory, i.e.~in classical mechanics, and also in Bohmian mechanics. But then, why not make the assumption of fundamental randomness, as a power of nature, an explicit assumption? Admittedly, for practical reasons it might be useful to keep standard mathematics for our computations. For, in any case, both classical mathematics and the mathematics adapted to finite information always agree on the outcomes of computations. But fundamentally, we have to choose a perspective. Either all digits of the initial conditions are assumed to be determined from the first moment, leading to timeless physics, or these digits are initially truly indeterminate and physics includes events that truly happen as time passes \cite{NGNaturePhysComment20}. Notice that in both perspectives chaotic systems would exhibit randomness. In the first case, from the point of view of classical mathematics, all the randomness is encoded in the initial condition. In the second case, randomness happens as time passes. This is the point of view of intuitionistic mathematics, as developed by L.E.J. Brouwer, where the dependence on time is essential \cite{StandfordEncyclodediaIntuitionism}.

Quite independently of the above motivations, L.E.J. Brouwer, a 20th century Dutch mathematician, refused to admit ungraspable actual infinities in the foundations of mathematics \cite{Brouwer1948}. Accordingly, for Brouwer and his followers, the mathematical continuum can't be an infinite collection of completed individual points, each represented by a real number and each being 3 times infinite: an infinite equivalence class of infinite series of rational numbers, each rational being itself an equivalence class of infinitely many pairs of natural numbers. Brouwer denied that the continuum is built up from independently existing points: there must be more to the continuum than the collection of (lawlike, that is, algorithmically generated) real numbers \cite{PosyBook}. Brouwer - the mathematician - also insisted that in some cases the future values (of numbers) must be indeterminate, and mathematical experience is a deeply non-deterministic temporality \cite{PosyBook}.

Reading Brouwer's intuition about the continuum with the eye of today's scientists makes it hard to understand (though see the very accessible book by Carl Posy \cite{PosyBook}). Indeed, today's scientists have been selected by their ability, among others, to digest modern mathematics. However, Brouwer was able to develop an alternative mathematics which avoids ungraspable infinities, specifically to avoid the standard description of the continuum as an infinite collection of individual completed real numbers. Moreover, as we shall see in the next section, Brouwer describes the continuum as a ``viscous" medium \cite{viscous}, as he - and I - desired. Brouwer's construction of the continuum is the heart of intuitionistic mathematics. 

Intuitionistic mathematics is rarely presented, and when it is, it is presented in all its complexity, including intuitionistic logic and axiomatics (about which not all intuitionists agree). I believe this is not necessary, at least not for physics. Moreover, there is an intimate connection between intuitionistic mathematics and indeterminism in physics. One may guess that intuitionistic mathematics is poorly known also because of Brouwer's complex personality, quite an extreme idealist, while applying his mathematics to physics should be the task of a solid realist, aiming at bringing physics closer to our experience.

The objective of this article is to allow physicists and philosophers of science (and anyone interested) to make their first steps in intuitionism and to understand why this allows one to re-enchant physics, introducing a model of an objective ``creative" time, i.e.~a dynamical time that allows for an open future and the passage of time. Much remains to be done; I don't claim completion of a program, but rather, more modestly, to open a new door. Indeed, changing the mathematical language used by physics, from classical Platonistic to intuitionistic mathematics, could well make it easier to express some concepts and to rebut Dolev’s claim that passage cannot be part of physics \cite{DolevSilence}.

\section{Intuitionistic Mathematics}
Mathematics is the language of nature, as famously claimed by Galileo and most scientists. But which mathematics? Classical mathematics, i.e.~God's  mathematics, in which every number is a completed individual number, although most of them are not computable, a mathematics that assumes omniscience, i.e.~that every proposition is either true or false. Or a form of constructive mathematics, i.e.~a human's mathematics concerned with finite beings, a mathematics that doesn't postulate the law of the excluded middle because humans can't prove every proposition true or false. Obviously, the sort of mathematics physicists use greatly influences, almost determines, what physics says about nature. It is not that these different mathematical languages change the everyday working of physics; the practical predictions are the same. However, the mathematical language strongly suggests the worldview offered by physics. From God's viewpoint, everything is already there, fully determined, waiting for us to discover, to sort out into categories and sets. But for humans, things are always unfinished, continually evolving, progressing; only the computable is fully determined and only the finite can be grasped.

Erret Bishop \cite{BishopConstructivism} wrote beautifully about the constructive mathematics side of our story (though without mentioning physics): {\it The classicist wishes to describe God's mathematics; the constructivist,
to describe the mathematics of finite beings, man's mathematics for short}; {\it how can there be numbers that are not computable (...)? Does that not contradict the very essence of the concept of numbers, which is concerned with computations?} and {\it Constructive mathematics does not postulate a pre-existent universe, with objects lying around waiting to be collected and grouped into sets, like shells on a beach}. It is worth also quoting Carl Posy \cite{PosyBook}: {\it We humans have finite memories, finite attention spans and finite lives. So we can fully grasp only finitely many finite sized pieces of a compound thing; There’s no infinite helicopter allowing us to survey the whole terrain or to tell how things will look at the end of time.}

In this section I would like to present a succinct introduction to a specific form of constructive mathematics:  intuitionistic mathematics. I'll emphasize the aspect of intuitionistic mathematics which is most relevant to physics, in particular to an indeterminsitic physical worldview. Accordingly, I emphasize in which sense numbers in intuitionism are processes that develop in time. However, contrary to Brouwer and Bishop, I do not think of these processes as driven by some {\it idealized mathematician} \cite{Brouwer1948} nor {\it constructing intelligence} \cite{BishopConstructivism}, but merely driven by randomness, a randomness considered as a power of nature, like other laws of nature. Hence, Brouwer, the father of intuitionistic mathematics, would probably not have liked my presentation, because in my view nature (physics) plays an essential role, while Brouwer was a strong idealist. Admittedly, here mathematics is presented as a tool to study nature, i.e.~for doing physics.

In my presentation I limit myself to the basic tool - sequences of {\it computable numbers} determined by random bits - and the two main consequences - the continuum is viscous and the non-validity of the law of the excluded middle. I believe that all this is quite easy to grasp intuitively. 

The most common case found in the intuitionistic literature defines {\it real numbers} as provably converging sequences of rational numbers. This is very similar to the standard view of real numbers. The main difference is that, in intuitionistic mathematics, at every instant only a finite initial sequence is determined, but the future of the sequence may still be open, i.e.~undetermined. As time passes, that initial sequence develops with the addition of new rational numbers that get determined by fresh information. Hence, from the very essence of intuitionist numbers, time is essential in intuitionistic mathematics \cite{StandfordEncyclodediaIntuitionism}. This strongly contrasts with numbers in classical mathematics where all objects, including real numbers, are considered to exist outside time, in some Platonistic world, i.e.~are considered from a God's eye point of view. Accordingly, it should not come as a surprise that classical mathematics makes it difficult to describe the passage of time in physics, while intuitionistic mathematics allows for such a description. Somehow, time is expelled from classical mathematics; hence also the sense of the flow of time is expelled from physics when it ``talks" the language of classical Platonistic mathematics.

Here however, we don't start with sequences of rational numbers, but use sequences of computable numbers, i.e.~numbers defined by a finite deterministic algorithm, a concept that was under development at Brouwer's time, but which is much better suited for applications in physics. Indeed, in physics one often uses computable functions, like exponentials and sines, to describe the evolution of simple dynamical systems. Such functions are expressed by algorithms that map rational and computable numbers to computable numbers\footnote{Note that there are some contrived counter-examples that build on discontinuity, see \cite{PourEl}}. Hence, if one would like to keep these simple descriptions, one had better concentrate on computable numbers.

We shall also relax some constraints, requiring only convergence of the sequences of computable numbers with probability one, not with certainty\footnote{Admittedly, introducing probabilities into these sequences may re-introduce the classical real numbers into intuitionism, something one would like to avoid. Actually, among all the examples presented in the next section, only the last one assumes convergence with probability one, see subsection \ref{AutonomousNumbers}. Here I leave this as an open question for future work, however see \cite{vanFrassen,MorganLebland}}. In this way we adapt the intuitionistic concept of number to the common usage of physicists. Indeed, physicists often use intuitionistic concepts and reasoning without realizing it. Common sense leads physicists to realize that some classical mathematical facts are physically meaningless, like the infinitely precise initial conditions necessary for deterministic chaos in classical mechanics and the finite probability that a billiard ball tunnels through a wall in quantutm mechanics. For example, Born, one of the giants of quantum theory, wrote: {\it Statements like `a quantity x has a completely definite value' (expressed by a real number and represented by a point in the mathematical continuum) seem to me to have no physical meaning} \cite{Born}; see also Drossel's view on the role of real numbers in statistical physics \cite{Drossel}. Also, scientists working on weather and climate physics explicitly use finite-truncated numbers and stochastic remainders \cite{PalmerStochClimateModel}. In a nutshell, for physicists, ``real numbers are not really real" \cite{NGrealNb}. It is only when real numbers and some other mathematical objects are taken literally that some odd conclusions follow, like - again - that classical mechanics imposes a deterministic worldview\footnote{Note that classical physics is compatible with a deterministic worldview; we merely stress that this is not the only possibility.}.

\section{Intuitionism: a first encounter}
Intuitionists reject ungraspable infinities. Hence, they reject the typical real numbers of classical mathematics\footnote{Emile Borel nicely illustrated the infinite amount of information contained in typical real numbers by noticing that the digits of one single real number could contain the answers to all questions one could formulate in any human language \cite{Borel}.}. For an intuitionist at any time-instant, every number is determined by only finite information, for example by a finite series of digits, perhaps generated according to some principle, but still only a finite number. However, this series of digits is not frozen, but evolves as time passes: new digits can be added, though only a finite number at a time. More precisely, new information is added - again perhaps according to some principle - but only a finite amount of information. Hence, let me stress that numbers are processes that develop as time passes. The new information can be entirely fresh. Brouwer, as a proper idealist, thought of a sort of idealized mathematician who would produce this new information, e.g.~by solving mathematical problems \cite{PosyBook}. This is not very appealing to physicists, nor more generally to realists. Nevertheless, the idea that the passage of time, the creation of new information and an indeterminate future may enter at such a basic level as numbers is highly attractive to those who believe that time, passage and an open future are essential features of reality.

Hence, let us posit that nature has the power to continually produce truly random bits. More precisely, at every discrete instant of time $n$, where $n$ denotes a positive integer, a fresh new random bit, denoted $r(n)$, comes into existence. It is totally independent of all the past, in particular of all previous random bits $r(1), r(2), ..., r(n-1)$. The most straightforward way to use these random bits to construct an intuitionistic real number $\alpha$ in the unit interval $[0..1]$ is to use binary format and to add this fresh bit to the already existing series of bits:
\beq
\alpha(n) = 0.r(1)r(2)r(3)...r(n)
\eeq
The above expression of $\alpha$ should be seen as a never ending process, a process that develops in time. Intuitively, as time passes the series $\alpha(n)$ converges to the real number\footnote{Admittedly, we use the same notation $\alpha$ for the sequence and for the real number it converges to, despite the fact that these are two different things, hoping the reader will not get confused by this.} $\alpha$.
Crucially, at any finite time, only finitely many bits $r(n)$ are determined and thus accessible.

This simple first example of intuitionistic real numbers cleanly emphasizes the intrinsic randomness of the bits in the binary expansion of all typical real numbers \cite{Chaitin1,Chaitin2}. 

The above expression for $\alpha$ is only the simplest and most straightforward intuitionistic element of the continuum. A first slight generalization assumes that the series of random bits starts at any position; that is, one could add any integer $\alpha_0$ to $\alpha$ and some initial finite series of bits $\lambda_0$:
\beq
\alpha(n) = \alpha_0.\lambda_0r(1)r(2)r(3)...r(n)
\eeq
More generally, intuitionist numbers allow one to use all the existing random bits in any (finite) computable way. 

\begin{figure}[h]
\includegraphics[width=4cm]{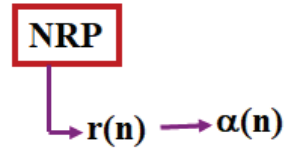}
\caption{\it We model an indeterministic world by a Natural Random Process (NRP) and by intuitionist numbers $\alpha$ that are processes fed by the NRP. At every time-instant $n$, the output of the NRP, denoted $r(n)$, is used to determine the computable number $\alpha(n)$. The sequence of $\alpha(n)$ converges, but, generally, at any finite time the sequence is still ongoing, hence the intuitionist number $\alpha$ is never fully determined. This illustrates the fact that at any finite time, $\alpha(n)$ contains only a finite amount of information, it is only the entire sequence of $\alpha(n)$ that may contain an unlimited amount of information. The collection of all $\alpha$'s recovers the continuum, though a sticky or viscous continuum, because at every time there are some sequences $\alpha(n)$ still ongoing. Hence, at any finite time, one can't pick out a point $\alpha$ out of the continuum, as it is not yet determined where exactly that sequence will converge. Somehow, it sticks to all the other $\alpha$'s that had, so far, the same sequence of $\alpha(n)$. This is in sharp contrast to the classical description of the continuum where every point is represented by a completed real number. Somehow, the classical continuum is analogous to the intuitionistic continuum, but viewed from the ``end of time", i.e.~a God's eye view.}\label{figNRP}
\end{figure}

Accordingly, as illustrated in Fig.~\ref{figNRP}, an intuitionist real number $\alpha$ is given by a sequence of computable numbers $\alpha(n)$ for all positive integers $n$ satisfying the following conditions:
\begin{enumerate}
\item $\alpha(0)$ is a given computable number.
\item For every integer $n\geq1$ there is a random bit $r(n)$ and all $r(n)$'s are independent of each other, i.e.~the random bits are assumed to be i.i.d.\footnote{Independent and identically distributed.} with uniform probability.
\item There is a computable function\footnote{My definition is close to what modern intuitionists call ``projections of lawless sequences", see p.68 of \cite{PosyBook}. The function $f\!ct$ acts in a way similar to what Brouwer called a ``spread", see p.30-31 of \cite{PosyBook}.}, denoted $f\!ct$, s.t.
\beq\label{fct}
\alpha(n)=f\!ct\big(\alpha(n-1),n,r(1),...,r(n)\big),
\eeq
hence, $\alpha(n)$ is a computable number for all $n$.
Note that the function $f\!ct$ doesn't need to depend on $\alpha(n-1)$, as $\alpha(n-1)$ is determined by the available information $\big(\alpha(0),n-1,r(1),...,r(n-1)\big)$; but it is simpler to define $f\!ct$ as in (\ref{fct}).
\item The sequence $\alpha(n)$ converges with unit probability (where the probability is over the random bits). For example one may impose $|\alpha(n)-\alpha(n-1)|\leq 2^{-n}$ which guarantees convergence.
\end{enumerate}

Brouwer named the series $\alpha(n)$ {\it choice sequences}. Here we use equivalently the terminology of sequence or series $\alpha(n)$.

Let me emphasize several important points:
\begin{enumerate}
\item Notice how general this definition is. Indeed, the function $f\!ct$ is only required to be computable, but beyond that can be whatever. As it depends on the time-instant $n$, it can be tailored to adapt to each instant $n$; it may fully depend on the values of the random numbers $r(n)$. Note that one does not need to restrict the $r(n)$'s to bits; they could as well designate any finite fresh information, e.g.~any finite series of random bits, as illustrated in subsection \ref{MortalNb}.

\item I like to think of the random bits as produced by a Natural Random Process: {\bf NRP}. The basic idea is that if indeterminism exists, then, as time passes, nature is able to produce true random bits $r(n)$, i.e.~true little acts of creation which definitively differentiate the past, the present and the future: at time-instant $n$, past random bits $r(p)$, $p<n$ are determined, $r(n)$ has just been created, while future $r(f)$, $f>n$ are indeterminate\footnote{If we want to be able to conceptualize and contemplate ``indeterminism" - and not stay stuck with determinism - we have to admit as a basic fact of nature that entirely new events happen, that new information gets created, for example in the form of new random bits, that the NRP considered here continually outputs.}. These random bits are used in a process that eventually determines a real number. Hence, numbers are processes. Note that at any finite time, the process is still ongoing, i.e.~the number is never fully determined, except for a countable subset of all real numbers. From the end of time point of view, the set of all intuitionist numbers $\alpha$ cover all the continuum. But that's only the end of time viewpoint. More precisely, it is impossible to describe classically a real number such that one knows (intutionistically or classically) that no intuitionistic real number will ever coincide with it.

\item Each sequence $\alpha(n)$ uses one NRP. A priori, two sequences $\alpha(n)$ and $\beta(n)$ use two independent NRP, although it would be interesting to also consider correlated NRP.

\item Obviously, intuitionistic mathematics requires time \cite{StandfordEncyclodediaIntuitionism}, just as counting requires time. This is not surprising; new information means that there was a time when this information did not exist, so indeterminism implies an open future. Time passes, we all know that, and intuitionistic mathematics integrates that fact in its heart and builds on this fact. This is in strong contrast to classical mathematics, which assumes numbers and other objects to be given all at once, somehow from outside time, or from the end of time, existing in some idealized Platonic world. Hence, unsurprisingly, classical mathematics is at odds with time and is a poor tool for theories that like to incorporate a description of time, or at least a description of stuff that evolves in time.

\item At every time-instant $n$, all future random bits $r(f)$, for $f>n$, are totally indeterminate. 

\item At every time-instant the process $\alpha(n)$ might still be ongoing, possibly forever. However, it is also possible that at some instant $n_d$ the processes terminates (dies) and all future $\alpha(f)$ equal $\alpha(n_d)$: $\alpha(f)=\alpha(n_d)$ for all $f\geq n_d$, see examples in subsection \ref{MortalNb}.

\item At any finite time-instant $n$, the theretofore determined sequences $\alpha(n)$ constitute only a countably infinite set. Indeed, for a given computable function $f\!ct$ there are only finitely many sequences $\alpha(n)$ determined by the finitely many $r(k)$, $k=1...n$. And there are countably infinitely many computable functions.

\item We define autonomous numbers as those sequences where the function $f\!ct$ does not depend on $n$. 

\end{enumerate}

The following sub-sections present examples of intuitionist numbers defined as choice sequences based on the Natural Random Process NRP. These examples assume binary notations and numbers in the unit interval $[0..1]$, but can easily be extended to arbitrary numbers in any basis.

\subsection{Totally Random Numbers}
\begin{figure}[h]
\includegraphics[width=6cm]{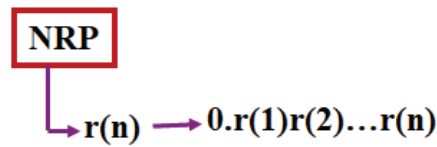}
\caption{\it Assume the NRP outputs independent random bits, $r(n)$, all uniformly distributed. At each time-instant $n$, the bit $r(n)$ is merely added to the bit series of $\alpha$, i.e.~$\alpha(n)=0.r(1)r(2)...r(n)$. Such choice sequences define typical real numbers, i.e.~numbers with no structure at all. We name such intuitionist numbers {\it Totally Random Numbers}}\label{figTotal}
\end{figure}

Assume the NRP outputs at each instant, $n$, one bit $r(n)$, see Fig.~\ref{figTotal}.

Define $\alpha(n)=\alpha(n-1)+r(n)\cdot 2^{-n}$. Hence:
\beq
\alpha(n)=0.r(1)r(2)...r(n)
\eeq
Here we assumed $\alpha(0)=0$, but one can easily allow any initial $\alpha(0)$ and merely add it to the above.

The conditions defining intuitionist numbers are clearly satisfied. We name such $\alpha$'s Totally Random Numbers. They correspond to typical real numbers, and all real numbers can be seen as the result of such a process, including the typical non-computable real numbers\footnote{Note that one may somewhat artificially construct ``mixed" numbers, e.g.~every second bit is random while every other bit is given by the corresponding bit of a computable numbers like, e.g., $\pi$.}. However, intuitionist numbers are much richer, as illustrated in the following sub-sections.

\subsection{Computable Numbers}\label{CN}
The computable function $f\!ct$ could be independent of all the random bits $r(n)$ and determine one computable number. A well-known example is the number $\pi$, the ratio of the circumference of a circle to its diameter.

Computable numbers are thus clear examples of intuitionist numbers, though not the typical ones. Here, there is nothing creative; there is no creative dynamical time, though computing the next digit, or next approximation of a computable number, necessarily takes some time. 

Note that the bits (and digits) of computable numbers may look random, even if one knows the function $f\!ct$, i.e.~one knows the algorithm that allows one to compute it up to any arbitrary precision. However, this apparent randomness differs profoundly from true randomness, randomness that involves the creation of new information. One way to illustrate this is a remarkable algorithm \cite{piDigits}
that allows one to compute the number $\pi$:
\beq\label{pi}
\pi=\sum_{k\geq0}\frac{1}{16^k}\big(\frac{4}{8k+1}-\frac{2}{8k+4}-\frac{1}{8k+5}-\frac{1}{8k+6}\big)
\eeq
There are many different algorithms (i.e.~different $f\!ct$'s) that compute $\pi$, but this one is of special interest here because it allows one to compute the bits of $\pi$ at any position without the need to first compute all the previous bits\footnote{The formula (\ref{pi}) is tailored for hexadigits, i.e.~digits in bases 16, but also applies to base two, i.e.~to bits. In a nutshell, the $n$th hexadecimal is given by $\left\lfloor\pi 16^n\right\rfloor$ modulo 16 (where $\left\lfloor x\right\rfloor$ denotes the closer integer smaller than x). Apply that to (\ref{pi}) and split the sum in one sum over $k\leq n$ and a sum over $k>n$. Realize that the second sum doesn't contribute, hence a finite computation suffices to compute the $n$th hexadecimal.}. This clearly illustrates that all bits of $\pi$ already exist and can be accessed, without the need to wait a time corresponding to their positions in the series of bits. This is in total opposition to the totally random numbers presented in the previous subsection. Consequently, the randomness of the bits of computable numbers is only apparent, as they are all fully determined by a finite algorithm\footnote{In an indeterministic world the weather in both one and two years' time is, today, undetermined. In two years time it will be determined. However, first the weather in one year from now will be determined. This is in strong contrast to the bits of $\pi$ that can be accessed - and are thus determined - without first accessing the previous ones.}.\\

Another interesting example of computable numbers are so called pseudo-random series of bits, as produced by all modern computers and heavily used in today's cryptography. There are many families of pseudo-random numbers, each defined by a computable function $f\!ct$. Each function uses a finite sequence of bits, $r(1)...r(k)$, for a fixed integer $k$, as a seed to generate highly complex sequences of bits, complex enough that without the knowledge of the seed it is extremely difficult, possibly impossible in practice, to guess the next bit from only the knowledge of the function $f\!ct$ and of the previous bits. For an intuitionist these pseudo-random series of bits are just an example of a choice sequences $\alpha(n)$ where the function $f\!ct$ depends only on the first $k$ random bits $r(1),r(2),...,r(k)$. In practice, our computers emulate the NRP with movements of the mouse or the coincidences between a key stroke and the computer's internal clock, or similar event arising from outside the computer. Clearly, after time-instant $k$, the random bits that define the seed are determined and the pseudo-random sequence is fully determined, as any computable number.

Accordingly, computable numbers contain finite information even in the limit $n$ going to infinity, i.e.~the information defining the algorithm plus possibly a finite set of random numbers that may determine the seed of pseudo-random numbers. Hence, no $\alpha(n)$ ever contains more than that finite information. Consequently, even intuitionistically one can think of computable numbers as given all at once, contrary to all non-computable numbers that all contain an infinite amount of information. Note however, that, except for rational numbers, not all the infinite set of bits is given (determined) at once; only the information needed to compute any finite set of bits is given\footnote{For a variety of indeterminacy in intuitionism see \cite{IndeterminateNumbersPosy}.} (more precisely, only the information needed to compute any approximation $\pm2^{-n}$ is given at once).

\subsection{Finite Information Quantities - FIQs}\label{FIQs}
A particularly physically relevant example of intuitionist number is the following.
Let $k\geq3$ be a fixed positive odd integer.  At each time-instant $n$ we keep the $k$ last random bits: $r(n-k+1)...r(n)$ and forget about all ``older" (previous) random bits. Define $\alpha(n)$ as $\alpha(n-1)$ to which one adds as $n$th bit the majority vote of these last $k$ random bits. Formally:

{\bf\huge.} If $\sum_{j=1}^k r(n-j+1)>k/2$, then $\alpha(n)=\alpha(n-1)+2^{-n}$, else
 
{\bf\huge.} if $\sum_{j=1}^k r(n-j+1)<k/2$, then $\alpha(n)=\alpha(n-1)$.\\

\begin{figure}[h]
\includegraphics[width=6cm]{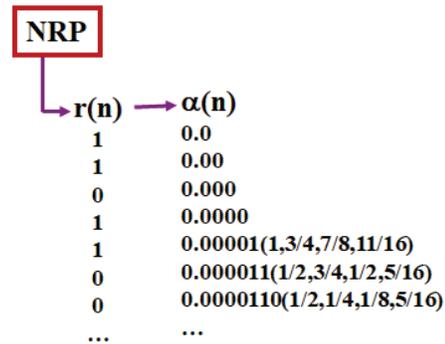}
\caption{\it Example of a Finite Information Quantity (or FIQ). At each time-instant $n$ the NRP outputs a bit $r(n)$. The $n$th bits of $\alpha$ is determined by the majority of the last 5 random bits $r(n-4)...r(n)$. At step $n$ the bits $n+1$ to $n+4$ of $\alpha$ are already biased either towards 1 or towards 0, indicated here in parenthesis by the propensities to eventually be determined (at later time-instants) by future outputs of the NRP to the bit value 1.}\label{FigFIQs}
\end{figure}

This corresponds to what we named ``Finite Information Quantities" (FIQs) in \cite{FlavioNG19}. At time-instant $n$, the $n$ first bits in the binary expansion of $\alpha$ are determined and equal either to 0 or to 1, and the far down the series of bits are totally indeterminate. However, interestingly, the $k$ bits in the intermediate positions, $n+1$ to $n+k-1$, may have a non-trivial ``propensity" to end up equal to 0 or to 1. For example, if the sum $\sum_{j=1}^k r(n-j+1)$ is larger than $\frac{k+1}{2}$, then the $(n+1)$th bit is already determined whatever $r(n+1)$ will turn out to be. Or, if the sum is smaller than  $\frac{k-1}{2}$, then the $(n+1)$th bit is already determined and equals 0. The next bit, i.e.~bit number $n+2$, might also be already determined, but there is a larger possibility that it is still undetermined, though not fully random: if the sum is rather large, then there is a similarly large ``propensity" that it will eventually get determinated to the value 1. The propensities of the further bits tend to move away from the extremal values 1 and 0 and from bit number $n+k$ all ``propensities" are fully random, i.e.~equal $\half$. Below and in Fig.~\ref{FigFIQs} an example is presented.

It is useful to introduce some notations. Denote $q_j$ the propensity of the jth bit to equal 1. Hence $q_j=1$ means that the jth bit equals 1 and similarly $q_j=0$ means that the jth bit equals 0. With this notation, at each time-instant $n$, $\alpha(n)$ can be expressed as a series of propensities, reminiscent of the usual series of bits, but where each $q_j$ is a rational number:
\beq
\alpha(n)=0.q_1...q_nq_{n+1}q_{n+k-1}\half...\half...
\eeq
where the first $n$ $q_j$'s necessarily equal 0 (if bit number $j$ is 0) or equals 1 (if bit number j is 1), while $q_{n+1}$ to $q_{n+k}-1$ equal rational numbers between 0 and 1 corresponding to the propensity that they eventually, after $k$ new time-instants, acquire the value 1 or 0 depending on future NRP outputs. 

Here is an example of the first steps of a FIQ (finite information quantity). Assume $k=5$ and the first 4 bits are given and all equal to 0, hence we set $\alpha(0)=\alpha(1)=...=\alpha(4)=0$. Assume the first 4 outputs of the NRP equal $1101$. Below we illustrate a possible growth of the FIQ sequence $\alpha$ using the propensity notations:
\beqa
r(1)...r(5)&=&\hspace{2.5mm} 11011 \\
\alpha(5)&=&0.000011\frac{3}{4}\frac{7}{8}\frac{11}{16}\half \label{q7} \\
r(2)...r(6)&=&\hspace{4mm} 10110 \\
\alpha(6)&=&0.000011\frac{1}{2}\frac{3}{4}\half\frac{5}{16}\half \\
r(3)...r(7)&=&\hspace{6mm} 01100 \\
\alpha(7)&=&0.0000110\half\frac{1}{4}\frac{1}{8}\frac{5}{16}\half
\eeqa

In this example, at time-instant $n=5$, the NRP outputs $r(5)=1$ which determines the 5th bit of $\alpha(5)$: $q_5=1$. Furthermore, the correlation present in this kind of random number is such that the 6th bit of $\alpha$ is already determined: whatever $r(6)$, one has $q_6=1$. Still at the time-instant $n=5$, the 7th bit is already biased towards 1: 3 out of the 4 possible random bits $r(6)$ and $r(7)$ determine that 7th bit to the value 1, hence, at that time-instant $n=5$, the propensity $q_7=\frac{3}{4}$, as indicated in (\ref{q7}). And so on, here assuming $r(6)=r(7)=0$. At time-instant $n$, at least $n$ bits of $\alpha$ are determined and at most $k-1$ bits are biased towards 0 or 1, i.e.~have a propensity different from $\half$.

It seems that all indeterministic physics can be modelled using only FIQs \cite{FlavioNG19}. Note that the FIQs presented here slightly differ from those of \cite{FlavioNG19} because here we may have correlations between the bits of $\alpha$. This allows one to define arithmetic for FIQs, see subsection \ref{arithmetic}.

Let us conclude this sub-section with a comment on the parameter $k$ that enters the definition of a FIQ. One may ask what the value of $k$ is for the dynamical system one is considering. Such questions are usually hidden in classical statistical mechanics, where one assumes that all initial conditions that satisfy some constraints, like fixed energy, are equi-probable. Here also, one could ask whether real-valued initial conditions with correlated bits, with correlation length $k$, are not more likely than others.

\subsection{Mortal Numbers}\label{MortalNb}
Here is another example of intuitionist numbers defined by a sequence much inspired by Posy's example based on the Goldbach conjecture, see \cite{Brouwer1948,IndeterminateNumbersPosy}. In this example the sequence may terminate (die) after a finite time or may continue forever, depending on chance, i.e.~depending on the outputs of the NRP. We name such numbers mortal, because the corresponding process may suddenly end.

Define $\alpha(1)=0.1$ and set a flag\footnote{Note that the value of this flag $f(n)$ can be computed from the series $r(1)...r(n)$, but it is simpler to keep track of it.} $f=1$. Assume that at time-instant $n$ the NRP outputs not just one bit, but outputs $n$ (independent) random bits $r(n)_1,r(n)_2..r(n)_n$.  

For all $n\geq2$:\\
{\bf\huge.} If the flag is set to 1 ($f=1$) and at least one of the $n$ random bits $\{r(n)_j\}_{j=1..n}$ equals 0, then $\alpha(n)_n=1$, i.e.~one adds a bit 1 at the end of the bits defining $\alpha(n)$ and the flag remains at 1.\\
{\bf\huge.} If the flag is set to 0 or if all random bits are equal to 1, then the series terminates, i.e.~$\alpha(n)=\alpha(n-1)$ and the flag is set to 0. 

Formally:
\beqa
f(n)&=&f(n-1)\cdot\big(1- r(n)_1\cdot r(n)_2\cdot...\cdot r(n)_n\big) \\
f(n)&=&0 \hspace{5mm}\Rightarrow\hspace{5mm} \alpha(n)=\alpha(n-1) \\
f(n)&=&1 \hspace{5mm}\Rightarrow\hspace{5mm} \alpha(n)=\alpha(n-1)+2^{-n}
\eeqa

Interestingly, since the probability that the flag is switched to 0 decreases exponentially with the time-instant $n$, there is a finite chance that the flag remains at 1 and the $\alpha(n)$'s increase continuously, approaching 1. The probabilities that the series of $\alpha(n)$ terminates at time-instant 
$n$ is $2^{-n}$. Hence, the probability that the series never dies equals:
\beqa
Prob(endless)&=&Prob(\alpha=1)\nonumber\\
&=&\Pi_{n\geq2}(1-2^{-n})\approx 0.5776
\eeqa

Consequently, a priori one doesn't know whether the series of $\alpha(n)$ will terminate and settle to a rational number $0.11..1$ with $k$ 1's, where $k$ is a priori undetermined, or whether the series converges to 1 (recall that $1=0.11..1..$, with an unending series of 1's). Hence, as long as the sequence $\alpha(n)$ hasn't terminated, it is undetermined whether $\alpha=1$ or $\alpha<1$, i.e.~$\alpha=1$ is not true and $\alpha\neq1$ is not true, an example of intuitionistic logic in which the law of the excluded middle doesn't hold. This should not surprise us, because the question whether $\alpha$ equals 1 or not is a question about the future, the open future! If one asks a question about the weather in a year's time, no one is surprised that the answer is not yet determined, i.e.~that the statement ``it will rain on the fieldhockey pitch in Geneva in precisely one year from now" is presently neither true nor false. At least, that's the case in an indeterministic world. As Posy put it \cite{PosyBook}: {\it The law of the excluded middle fails because objects and the world are not-determinate, so truths about them are indeterminate}.\\

Here is another example of a mortal number which might oscillate forever between below and above $\half$. The rule for the flag is as above and if $f(n)=0$, then $\alpha(n)=\alpha(n-1)$, again as above. However, if $f(n)=1$, define $\alpha(n)=\half+(-2)^{-n}$, for all $n=2,3,4,...$. Accordingly, as long as the sequence doesn't terminate, $\alpha$ oscillates between below and above $\half$: $\half,\half+\frac{1}{4},\half-\frac{1}{8},\half+\frac{1}{16},\half-\frac{1}{32},...$. This mortal number illustrates the concept of {\bf viscosity}: there is no way to cut the continuum in two, as there always are intuitionist numbers which are never definitively on one side of the cut, nor definitively on the other side, as here illustrated with a tentative cut at $\half$. Note that a straightforward consequence is the absence of step functions and more generally of discontinuous total functions of the unit interval\footnote{A function is total if it is defined everywhere.}, as it would be impossible to determine the value of the function at the point(s) of discontinuity. \\

Yet another mortal number is defined as follows. Let $r(n)=\pm1$ and define $\alpha(n)=\half+r(n)\cdot10^{-n}$. The series terminates at time-instant $n$ when, by chance,  the previous $n/2$ random bits $r(j)$ happen to have the same value, all +1 or all -1, and $n$ is even and larger or equal to 4. Clearly, for $n=4$ the series terminates whenever $r(3)=r(4)$, i.e.~with a probability of $\half$: $P_{stop}(n=4)=\half$. For $n=6$ the series terminates whenever $r(6)=r(5)=r(4)\neq r(3)$, where the last constraint implies that the series did not already terminate at instant $n=4$: $P_{stop}(n=6)=1/8$. However, computing $P_{stop}(n)$ for arbitrary even $n$ is non-trivial because for $n\geq8$ the 4 last random bits do not overlap with the early bits. The general formula reads, for all $n\geq3$:
\beq\label{Pstop}
P_{stop}(2n)=\big(1-\sum_{j=2}^{\lfloor n/2\rfloor} P_{stop}(2j)\big) \cdot 2^{-n}
\eeq
where $\lfloor n/2\rfloor$ denotes the largest integer smaller or equal to $n/2$. The first term in (\ref{Pstop}) is the probability that the series did not stop before instant $2\lfloor n/2\rfloor$ and the factor $2^{-n}$ the probability that the last half of bits are all equal and the series did not stop after instant $2\lfloor n/2\rfloor$ but before instant $2n$.

Accordingly, the probability that the series terminates reads:
\beqa
P_{stop}&=&P_{stop}(4) + \sum_{n\geq3} P_{stop}(2n) \\
&=&\half + \sum_{n\geq3} 2^{-n} - \sum_{n\geq3}\sum_{j=2}^{\lfloor n/2\rfloor} P_{stop}(2j) 2^{-n} \\
&=&\frac{3}{4} - \sum_{n\geq3}\sum_{j=2}^{\lfloor n/2\rfloor} P_{stop}(2j) 2^{-n} \\
&<&\frac{3}{4}
\eeqa
where we used $\sum_{n\geq k}2^{-n}=2^{1-k}$. Accordingly, the probability that the series never terminates is larger than a quarter.
Numerically I found $P_{endless}\approx 0.31668$. Importantly, this number is strictly positive.\\

It is fun to invent further mortal numbers. Appendix \ref{MortalFurmulas} presents some useful general formulas to evaluate the dying probabilities of mortal numbers.

Finally, no autonomous number can die with certainty. However, the next subsection shows that autonomous numbers can die with probability one.

\subsection{Autonomous numbers}\label{AutonomousNumbers}
So far all non-computable numbers we presented use a function that depends explicitly on the time-instant $n$, i.e.~are not autonomous. It is difficult to see how to guarantee convergence without such a dependence. However, it is natural to ask for autonomous numbers, i.e.~numbers defined by a function independent of $n$. Here is such an example:
\beq\label{eta}
\alpha(n+1) = \alpha(n) + r(n)\cdot\eta\cdot\alpha(n)\cdot\big(1-\alpha(n)\big)
\eeq
where $r(n)=\pm1$ denotes the outputs of the NRP and $\eta$ is any given computable number between 0 and 1 which determines the rate of convergence. 

This sequence has two fixed points at 0 and 1. Interestingly, for any value of $\alpha(n)$ the probability (over the $r(k)$ for $k>n$) that the sequence converges to 1 equals $\alpha(n)$. This sequence converges only with probability one, as some sequences of random numbers $r(n)$ do not lead to convergence, like an infinite sequence of alternating\footnote{More generally, let $r(n)=sign[\half-\alpha(n-1)]$.} $+1$ and $-1$ and $\alpha(0)=\frac{\eta+2-\sqrt{\eta^2+4}}{2\eta}$.

This example is inspired by the old gambler's ruin problem \cite{GamblerRuin}. Interestingly, this has inspired (sometimes unknowingly) early attempts to solve the quantum measurement problem by adding stochastic terms to the Schr\"odinger equation \cite{Pearle82,Gisin84}. This led to spontaneous collapse models, quite reminiscent of intuitionist numbers.

\subsection{Arithmetic and logic of intuitionist numbers}\label{arithmetic}
Numbers are there to count and more generally to compute and to compare. Intuitionist numbers can straightforwardly be used in any algorithmic computations. It suffices to apply the algorithm to each term of the sequence $\alpha(n)$. For example, the addition $\alpha+\beta$ is given by the sequence $\alpha(n)+\beta(n)$, and the exponential of $\alpha$ is given by the sequence $\exp\{\alpha(n)\}$. Since we defined intuitionist numbers as sequences of computable numbers (and not merely rational numbers), every computable function (i.e.~algorithmic function) can be applied to any intuitionist number, though care has to be paid to the convergence as, e.g., in the sequence $1/\alpha(n)$ that diverges if $\alpha=0$.\\

The set of intuitionist numbers can be ordered (see section 2.2.2 of \cite{PosyBook}):\\
$\alpha > \beta$ iff one can construct two integers $k$ and $n$ such that for all $m\geq0$
\beq\label{ordering}
\alpha(n+m)-\beta(n+m) \geq 2^{-k}
\eeq
At first sight this may look very similar to well-known definitions in classical mathematics. However, there are serious caveats:
\begin{enumerate}
\item First, the integers $k$ and $n$ have to be constructable, not merely exist in some Platonistic sense.
\item Second, these constructions should be possible at the time-instant $n_{now}$ when the assertion is made, i.e.~the constructions cannot access any of the future outputs $r(f)$ of the NRP, $f>n_{now}$, because these future $r(f)$ are not yet determined.
\end{enumerate}
Consequently, the above defines only a partial order, as the following does not hold for all $\alpha$ and $\beta$:
\beq\label{TotalOrder}
(\alpha=\beta) \vee (\alpha < \beta) \vee (\alpha > \beta)
\eeq
where $\alpha=\beta$ means that both sequences converge to the same real number whatever future bits are produced by the NRPs. A first example is provided by the mortal number $\alpha$ that oscillates between below and above $\beta=\half$ as long as it didn't die, see subsection \ref{MortalNb}. As another example, consider two FIQs, $\alpha$ and $\beta$, based on two independent NRPs, and assume that up to the present time-instant $n_{now}$ all the bits determined by then coincide: $\alpha(n)=\beta(n)$ for all $n\leq n_{now}$. That does not guarantee that at the next time-instant the next bits still coincide. But even if they do not coincide, e.g.~$\alpha(n_{now}+1)=1$ and $\beta(n_{now}+1)=0$, it could still be the case that the two sequences converge to the same value if, by chance, all future bits of $\alpha$ are 0's and all next bits of $\beta$ are 1's. All this may seem surprisingly, but only if one forgets that intuitionist numbers are processes that develop as time passes\footnote{Of course, at any time-instant, $n$, a version of (\ref{TotalOrder}) (where the ordering is over the computable numbers), holds for the computable approximations, $\alpha(n)$ and $\beta(n)$, of $\alpha$ and $\beta$ that are determined at $n$. However, for all $n$, there are $(\alpha,\beta)$ such that (\ref{TotalOrder}) does not hold at time-instant $n$ where the ordering is defined by (\ref{ordering}).}.

Here, again, it would be only at the end of time – if there were such a thing– that (\ref{TotalOrder}) could hold: at any finite time we have no ``infinite helicopter", using Posy's illustration, to see how the sequences will develop.\\

Let us consider one more definition. We say that $\alpha$ and $\beta$ are apart, denoted $\alpha\#\beta$, iff one can construct two integers $k$ and $n$ such that for all $m\geq0$
\beq
|\alpha(n+m)-\beta(n+m)|\geq 2^{-k}
\eeq
Intuitively, $\alpha$ and $\beta$ are apart if one can put one's finger in between $\alpha$ and $\beta$.
Interestingly, one can prove that if two intuitionist numbers are not apart, i.e.~if one can constructively prove that one can't put one's finger in between, then they are equal \cite{PosyBook}:
\beq
\neg(\alpha\#\beta) \Rightarrow (\alpha=\beta)
\eeq
From the above one is temped to deduce $(\alpha\neq\beta) \Rightarrow (\alpha\#\beta)$. But the latter is wrong! The assumption $\alpha\neq\beta$ does not allow one to construct the two integers $k$ and $n$ needed to prove ``apartness", a nice illustration of the non-validity of the law of the excluded middle in intuitionistic mathematics.

The fact that the law of the excluded middle fails in intuitionism is surprising to classical eyes. But, if one thinks more about this, it is very natural and even necessary in an indeterministic world. Indeed, at any finite time, there are propositions that can't be proven in a finite number of steps using only the information existing at that time. As previously stated, the excluded middle fails because the world is not-determinate, so truths about it are indeterminate. And - my emphasize - this holds also for mathematical objects.

A basic logical consequence runs as follows.
Assume that a proposition $P$ is true, i.e. there is a finite proof of $P$. Then, it is impossible to prove $P$ false, hence $P\Rightarrow \neg\neg P$. However, because of the lack of the excluded middle, a proof that it is impossible to prove $P$ false is not a proof of $P$: $\neg\neg P \centernot\Rightarrow P$. In physical terms: the impossibility to prove that it will rain in a year's time is no evidence that it will be sunny. Indeed, the weather in a year's time could merely be undetermined. However, a proof of the impossibility to prove that it will not be not rainy in a year time implies that it will not be rainy: $\neg\neg\neg P \Leftrightarrow \neg P$.

Because of the non-validity of the law of the excluded middle, some classical theorems are not valid intuitionistically. However, there are also some new theorems, invalid classically but valid intuitionistically, and some theorems valid both classically and intuitionistically, but that require very different proofs. Let me illustrate these 3 cases.

First, let's consider a theorem that holds in intuitionism, but not classically: all total functions (i.e.~functions defined everywhere) are continuous. This is known as Brouwer's theorem. This excludes, among others, step functions from the collection of total functions. Indeed, according to intuitionism, at no finite time could one define the value of the function at the point(s) of discontinuity, because, at that time, some choice sequences are still fluctuating above and below that point. Note, however, that intuitionism accepts arbitrarily close approximations to discontinuous functions, i.e.~functions with arbitrarily fast transitions from one value to another.

Second, here is an example of a classical theorem not valid intuitionistically. The classical intermediate value theorem is not valid in intuitionism: for a continuous function $f(x)$ for which there exist real numbers $a<b$ such that $f(a)<0<f(b)$, one can't construct a point $x_0$ s.t. $a<x_0<b$ and $f(x_0)=0$. At first sight this may look shocking to classical eyes. However, what physicists really need in practice is a weaker form of the intermediate value theorem, a form that holds intuitionistically: under the same assumptions and for all $\epsilon>0$ one can construct an $x_0$ s.t. $a<x_0<b$ and $|f(x_0)|<\epsilon$.

Finally, there are also theorems that hold in both classical and intuitionistic mathematics, but which require quite different proofs. An example is Gleason's theorem, which plays a central role in the foundations of quantum theory. Hellman noticed that the original proof by Gleason is not constructive, hence in particular not valid intuitionistically \cite{Hellman}. However, a few years later, Richman and Bridges gave a very different constructive proof of the original Gleason theorem \cite{RichmanBridges}.

In summary, there are deep differences between classical and intuitionistic mathematics. These difference are precisely those needed to describe indeterminacy and indeterminism, both in the physical world and in mathematics. What is important for the practitioner is that all of physics that can be simulated on a (classical\footnote{Some quantum information processing - on so-called quantum computers - cannot be efficiently simulated on classical computers. However, all quantum information processing can be simulated in a finite time on a classical computer. Accordingly, everything that can be simulated by a ``quantum computer" also holds intuitionistically.}) computer can also be derived using only intuitionistic mathematics.

\section{Indeterministic physics and intuitionistic mathematics}
If one wants to seriously consider the possibility of indeterminism, i.e.~to negate the necessity of determinism, it makes plenty of sense to postulate that nature is able to continually produce new information, here modeled by true random numbers. Indeed, the very meaning of indeterminism is that nature is able of true little acts of pure creation: the $n$th output of the NRP, $r(n)$, was not necessary before it happens, hence was totally impossible to predict, but after time-instant $n$ it is a fact of nature. These random numbers allow one to define choice sequences (in Brouwer's terminology) that represent the continuum, but with numbers that are not all given at once, contrary to Platonistic/standard mathematics, numbers that are processes that develop as time passes. Table \ref{table1} illustrates the close connection between concepts in indeterministic physics and in intuitionistic mathematics.\\

\begin{table}[ht]
	\centering
		\begin{tabular}
			{|c|c|c|}
			\hline			
			& \large Indeterministic & \large Intuitionistic \\ & \large Physics & \large Mathematics\\
			\hline
1 & Past, present and future	& Real numbers \\
 & are {\bf not} all given at once & are {\bf not} all given at once\\
			\hline
		2 &	Time passes	& Numbers are processes \\
			\hline
		3 &	Indeterminacy &	Numbers can contain only \\&& finite information\\
			\hline
4 &Experiencing	& Intuitionism rests on \\&& grasping objects\\
			\hline
5 &The present is thick	& The continuum is viscous \\
			\hline
6 & Becoming &	Choice sequences\\
			\hline
7 & The future is open &	No law of the excluded middle \\&&(a proposition about the future \\&&can be neither true nor false)\\
			\hline
		\end{tabular}
\caption{\it This table illustrates the close connections between the physicist's intuition about indeterminism in nature and the mathematics of intuitionism.} \label{table1}
\end{table}

Below I briefly comment each line of Table \ref{table1}.

\begin{enumerate}
\item In indeterministic physics the past, present and future are not all given at once, contrary to the block universe view. Analogously, in intuitionistic mathematics the continuum is described by numbers that are not all given at once; for most of them their series of digits is still an ongoing process, contrary to classical real numbers, whose infinite series of decimals is assumed to be completed since ever and for ever.

\item In indeterministic physics, time is modeled as passing, contrary to the block universe in which everything is fixed and frozen. Analogously, in intuitionistic mathematics the numbers that fill the continuum are processes that develop as time passes.

\item In intuitionistic mathematics, at any time, numbers, like all mathematical objects, are finite, in particular they contain finite information. Hence, if the complexity of the evolution requires unbounded information, as in classical chaotic dynamical systems, then the evolution is necessarily indeterministic.

\item Physics is not only about fascinating technologies and highly abstract and sophisticated theories. Physics should tell stories about how the world is and functions. Physics should help us to develop our intuition, like, e.g., how the kangaroos manage not to fall off Earth, how the moon drives tides, how transistors allow our computers to operate and how time passes. In a nutshell, physics should not be too far from our experiences. Similarly, mathematics should not be too far from our intuition and in particular should avoid the ungraspable infinities that plague classical Platonistic mathematics.

\item Many physicists have the intuition that the present is thick, that it can't be of measure zero, infinitely squeezed between the past and the future. The present and passage are necessary ingredients to tell stories. As Yuval Dolev nicely put it, ``To think of an event is to think of something in time" \cite{DolevSilence}. In intuitionistic mathematics the continuum is viscous; it can't be neatly cut in two. This might well provide the thickness that our model of the present needs for a faithful description.

\item Some things merely are, but most things are changing, events are becoming. Becoming is central in intuitionistic choice sequences, in the way bits come into existence one after the other.

\item In indeterministic physics the future is open, again is sharp contrast to the block universe view. Consequently, statements and propositions about the future need not be either true or false. For example, the proposition ``it will rain in exactly one year time from now at Piccadilly Circus" is neither true (because it is not predetermined that it will rain), nor is it false (because it is not predetermined that it will not rain). Hence, in a world with an open future, the law of the excluded middle does not always hold. Analogously, in intuitionistic mathematics, as long as a number is not completed, there are statements about it that are neither true nor false, and in intuitionistic logic the law of the excluded middle doesn't hold.
\end{enumerate}

\section{Intuitionistic and/or classical mathematics?}
There are several mathematical languages. It is not that one is correct and the other one wrong. Hence, the question is not intuitionistic or classical mathematics. Both exist on their own, independently of physics; both have their beauties and roles. However, the different languages make it clear that some conclusions one is tempted to infer from physics are, actually, inspired by the language, not by the facts. A central claim of this paper is that intuitionistic mathematics is better suited to describe a world full of indeterminacy, a world in which time passes and the future is open. However, admittedly, intuitionism does not prove that our world is indeterministic, it only proves that physics is equally compatible with an indeterministic worldview as it is with a deterministic one.


Intuitionistic mathematics is a form of constructive mathematics, i.e. all objects are defined by finite information at all times. Additionally, intuitionism incorporates a dynamical time. As we have seen, in intuitionistic mathematics, at every time instant, there is only finite information and there are ever ongoing processes; this is the mentioned dynamical time. I like to call this dynamical time ``creative time", as truly new information is continually created \cite{GisinTimePasses}. This new information feeds into the mathematical objects and is necessary to provide a mathematical framework to describe indeterminism in physics and the passage of time: there is a time before and a time after the creation of the new bits of information. Contrary to Bergson, I do see this dynamical/creative time as entirely objective, it is a purely natural process. One consequence is that the law of the excluded middle and the principle of sufficient reason don't hold in intuitionism, as, I believe, it has to be the case in any theory that faithfully describes the world as indeterminate and its evolution as indeterministic.
\\


\section{conclusion}
Physicists produce models of reality. The models should be as faithful as possible, in particular produce correct empirical predictions. This is the first criterion to judge physical models. However, it is not the only one. Physical models should also allow humans to tell stories about how nature does it, e.g.~how the moon drives the tides, how white bears and kangaroos remain on Earth, how lasers operate and how time passes. One should not confuse the model with reality, hence our models can at best help us to gain understanding and develop our intuitions of how nature does it. From this point of view, physics should model the passage of time. In this article we argued that this can be done by modeling dynamical ``creative" time at the level of numbers, by the continual creation of new information, modeled by new independent random bits. This should go down all the way into the mathematical language we use to formulate our physical models. Surprisingly, such a language has already existed for a century, with Brouwer's intuitionism and his choice sequences, and especially with Kreisel's lawless choice sequences \cite{KreiselLLCS}

In intuitionism, the law of the excluded middle holds only if one assumes a look from the ``end of time", that is, a God's eye view \cite{NGNaturePhysComment20}. But at finite times, intuitionism states that the law of the excluded middle is not necessary, that there are propositions that are neither true nor false, but merely undetermined. Such propositions might be about the future, the open future, as already emphasized by Aristotle \cite{Aristotle}. But they might also be about numbers, the numbers that are at the basis of the scientific language. In classical mathematics, these numbers are called ``real", for historical reasons. For intuitionists, those ``real" numbers are never completed, at least never in a finite time, which are the only times that there are. Their infinite number of digits, coding infinite information, are, again, only the view from the end of time. Accordingly, ``real" numbers hide all the future far down in their series of digits: ``real" numbers are the hidden variables of classical physics \cite{NGHiddenReals}. It is the common usage of real numbers in physics that produces the illusion that the future is already fixed. In a nutshell, ``real numbers are not really real" \cite{NGrealNb}, a fact deeply incorporated in intuitionistic mathematics.

Finally, replacing the real numbers physicists use with ``random" numbers, i.e.~intuitionist numbers based on natural random bits, as presented here, might turn out to help overcoming the conundrum in which today's physics is locked, between a quantum theory full of potentialities and indeterminacy and the block universe view provided by general relativity. Indeed, it is the Platonistic mathematics that physicists use unconsciously that leads them to trust the block universe view.


\small
\section*{Acknowledgment} 
Useful critics and comments by Carl Posy, Yuval Dovel, Gilles Brassard, Ben Feddersen, Barbara Drossel, Valerio Scarani, Christian W\"uthrich, Flavio Del Santo, Jon Lenchner, Tein Van der Lugt, Stefan Wolf and Michael Esfeld are ackowledged, as well as the many colleagues who send me comments on my Nature Physics contribution \cite{NGNaturePhysComment20}.
Financial support by the Swiss NCCR-SwissMAP is greatfully acknowledged.

\normalsize
\appendix\section{Quantum indeterminacy}\label{Qindeterminacy}
Quantum physics is ``officially" non-deterministic, following the standard Copenhagen interpretation. However, most physicists seem not to entirely admit that state of affairs. In particular, high-energy physicists mostly swear only by relativity and the (beautiful in their eyes) block universe, in which everything is frozen since ever and for ever. Physicists and philosophers working on the foundations of quantum physics also mostly adhere to alternative interpretations. Hence, in this appendix, I summarize my understanding of quantum indeterminacy.

Surprisingly, the official Copenhagen interpretation, with its top-down indeterministic causation from large classical measurement apparata down to microscopic quantum systems, is rarely seriously defended by physicists because of its vagueness, a weakness that admittedly plagues this standard interpretation. Hence, only spontaneous dynamical collapse models aim at providing a serious physical description of the indeterminism present in quantum physics \cite{GRW,belljumps,NGHPA89,RMPcollapseModels}. These models assume a fundamental random process that presents the Schr\"odinger equation as an approximation to a more fundamental stochastic evolution equation. The stochastic term in the equation is always present and active, though its effect is small for systems with a few particles and huge for system composed of many (Avogadro number) particles. The random process, typically a Wiener process, is presented as fundamental, not decomposable into more basic elements. It is not a field; it is merely a power of nature (a disposition) that affects all quantum evolutions.

The alleged incompleteness of quantum theory goes back to the hugely influential (and beautiful) paper by Einstein, Podolsky and Rosen in 1935 \cite{EPR35}. It sparked large research programs, which, on one side, definitively excluded the possibility of a completion in terms of local variables \cite{Bell64,GisinQchance14}, and on the other side led to Bohmian mechanics \cite{Bohm52,BellBohm,DurrTeufel}. Interestingly, the first side brought along the understanding that quantum indeterminism can be mathematically proven from two natural assumptions: no arbitrarily fast influences and no superdeterminism, plus the experimental fact that Bell inequalities can be violated \cite{Pironio10,Acin16,BrunnerRMP14,GisinQchance14}.

The second consequence of the EPR paper, i.e. Bohm's model, can be intuitively understand as follows. In Bohmian mechanics one adds positions of every particle to the usual quantum state and adds evolution equations of those particles' positions to the Schr\"odinger equation. All these evolution equations are deterministic and the apparent indeterminism is all due to the unknown initial particles' positions. As Bohmian mechanics' predictions are equivalent to standard quantum theory, it has to incorporate some non-locality. And indeed, the particles' positions evolutions depend not only on the quantum state nearby, but on the entire quantum state, including arbitrarily distant parts of the quantum wave-function. Bohmian mechanics is a perfect example that all indeterministic theoretical models can be complemented such that the complemented model is deterministic. In full generality, it suffices to assume as supplementary variables all results of all possible future measurements, while making sure that these remain inaccessible until the measurements take place \cite{NGrealNb}. In the case of Bohmian mechanics, this is done by assuming that the particles' positions remain hidden, i.e.~that one can't gain more information about them than what is provided by the usual quantum state and measurement results. This ``trick" guarantees also that the non-locality intrinsic to Bohmian mechanics can't be used for faster than light (actually, for arbitrarily fast) communication.

The many-world interpretation of quantum theory is another example of quite an extreme view that aims at circumventing quantum indeterminism \cite{Kent09}. Here, the basic idea is that all possibilities, however small their quantum probabilities, are realized in parallel. However, physicists don't realize this directly, because they also are in ``superposition" of experiencing all possibilities. And the physicists' friends also ``see" their friends with different experience, i.e.~the friends and eventually the entire universe evolve into an enormous superposition state in which everything that could happen actually happens \cite{WignerFriend67,RennerWigner18}. The evolution of this multiverse is described by the deterministic Schr\"odinger equation. Hence, indeterminism is expelled from this view, at the cost of an enormoulsy complex multiverse. Apparent indeterminism is merely due to our ``splitting" (superpositions) into several persons with different experiences.

\section{General formulas for mortal numbers}\label{MortalFurmulas}
Denote $q(n)$ the probability that a mortal series terminates at instant $n$, given that it did not stop before:
\beq
q(n)=Prob(stop ~@~ n|did~ not~ stop~ before)
\eeq
and $p(n)$ the probability that the series stops at instant $n$ and did not stop before:
\beq
p(n)=Prob(stop ~@~ n~ \&~ did~ not~ stop~ before)
\eeq
We get, for all $n\geq2$:
\beqa
p(n)&=& q(n)\cdot\big(1-\sum_{m=1}^{n-1}p(m)\big) \label{pnSum}\\
&=& q(n)\cdot\Pi_{m=1}^{n-1}(1-q(m)) \label{pnProd}
\eeqa

The equality between (\ref{pnSum}) and (\ref{pnProd}) is easy to prove by induction, starting from $p(1)=q(1)$. Hence:
\beq
P_{endless}=1-\sum_{n\geq1}p(n)=\Pi_{m\geq1}(1-q(m))
\eeq
Note that if $q(n)=2^{-n-1}$, then $p(n)\approx 30\%$, and if $p(n)=2^{-n-1}$ then $q(n)=\frac{1}{2+2^n}$.

There are not many infinite products with simple formulas. Consider the case $q(n)=1/n^2$, for all $n\geq2$, with $q(1)=0$. For example, at each time instant $r(n)$ consists of 2 n-dits, $r(n)=r_1(n),r_2(n)$ with $r_j(n)\in\{0,1,...,n-1\}$ and the series terminates iff $r_1(n)=r_2(n)=0$. Then $P_{endless}=\Pi_{n\geq2}\big(1-\frac{1}{n^2}\big)=\half$.

\end{document}